\documentclass{article}

\usepackage[final]{neurips_2024}
\usepackage{graphicx}

\usepackage[utf8]{inputenc} 
\usepackage[T1]{fontenc}    
\usepackage{hyperref}       
\usepackage{url}            
\usepackage{booktabs}       
\usepackage{amsfonts}       
\usepackage{nicefrac}       
\usepackage{microtype}      
\usepackage{xcolor}         

\usepackage{amsmath}
\title{Style Mixture of Experts for Expressive Text-To-Speech Synthesis}

\usepackage{authblk}
\author[1]{Ahad Jawaid}
\author[1]{Shreeram Suresh Chandra}
\author[2]{Junchen Lu}
\author[1]{Berrak Sisman}

\affil[1]{The University of Texas at Dallas, USA}
\affil[2]{National University of Singapore, Singapore}

\affil[ ]{\texttt{ahad.jawaid@utdallas.edu, shreeramsuresh.chandra@utdallas.edu} }
\begin{document}
\maketitle

\begin{abstract}
Recent advances in style transfer text-to-speech (TTS) have improved the expressiveness of synthesized speech. However, encoding stylistic information (e.g., timbre, emotion, and prosody) from diverse and unseen reference speech remains a challenge. This paper introduces StyleMoE, an approach that addresses the issue of learning averaged style representations in the style encoder by creating style experts that learn from subsets of data. The proposed method replaces the style encoder in a TTS framework with a Mixture of Experts (MoE) layer. The style experts specialize by learning from subsets of reference speech routed to them by the gating network, enabling them to handle different aspects of the style space. As a result, StyleMoE improves the style coverage of the style encoder for style transfer TTS. Our experiments, both objective and subjective, demonstrate improved style transfer for diverse and unseen reference speech. The proposed method enhances the performance of existing state-of-the-art style transfer TTS models and represents the first study of style MoE in TTS\footnote{Speech Samples: https://stylemoe.github.io/styleMoE/}.
\footnote{This work was funded by NSF CAREER award IIS-2338979.}
\end{abstract}

\section{Introduction}
Text-to-speech (TTS) frameworks have significantly evolved, aiming to produce speech that is not only intelligible but also rich in emotional and prosodic information, mirroring human-like expressiveness \cite{triantafyllopoulos2023overview}. Advances in neural TTS models have made significant improvements in generating high-fidelity speech~\cite{oord2016wavenet, wang2017tacotron, li2019neural}. 
The \textit{one-to-many} mapping issue in TTS presents a significant challenge, where a single text input can yield multiple speech outputs with a variety of speaking styles, depending on context, emotion, or speaker intention. Hence, traditional methods, such as
modeling under the L1 loss, often result in monotonous-sounding speech \cite{tan2021survey, 7393764}. To address this issue, researchers have developed various conditioning techniques. Initially, these techniques involved incorporating emotional and speaker identity labels \cite{7953089, lee2017emotional}. Subsequently, advancements led to direct style transfer from reference speech through neural style encoders to better capture expressiveness and prosodic information
\cite{skerry2018towards, wang2018style, akuzawa2018expressive}. Further refinements in modeling styles included integrating additional acoustic features ~\cite{ren2020fastspeech, lancucki2021fastpitch} and separating content from style for disentangled learning, and other strategies such as hierarchical encoding to capture stylistic details at different resolutions~\cite{sun2020fully, huang2022generspeech}.

Mixture of Experts (MoE) \cite{jacobs1991adaptive} is an ensemble technique that divides a complex problem space into more manageable subspaces, each handled by specialized experts. The gating mechanism facilitates the division by routing input samples to the most suitable expert(s), effectively implementing a "divide and conquer" strategy~\cite{jacobs1991adaptive, masoudnia2014mixture}. The MoE technique has been shown to enhance model generalization by favoring combining strategies that lower the variance error~\cite{masoudnia2014mixture}. Additionally, it can be used for conditional computation, allowing models to scale effectively without incurring proportional increases in computational costs~\cite{shazeer2017outrageously}. MoE has found significant applications in scaling large language models~\cite{shazeer2017outrageously} and vision models~\cite{riquelme2021scaling}, and in learning factored representations of datasets~\cite{eigen2013learning}, showcasing its versatility and efficacy across various domains.

Motivated by the capabilities of MoE, we propose StyleMoE, a novel approach to enhance style transfer in TTS systems, addressing the challenges of capturing diverse and unseen speaking styles. The proposed method replaces the style encoder in a TTS framework with a mixture of style experts alongside a gating network to select which expert to use based on the reference speech. To ensure this method maintains a computational cost similar to the original TTS model, we use a sparse MoE layer to limit the number of experts utilized during inference ~\cite{shazeer2017outrageously}. The proposed method allows for better style transfer from out-of-distribution reference speech since each style expert optimizes on a subset of data, which avoid the issue of learning averaged representations when optimizing a single style encoder. The main contributions of the paper are summarized as follows:
1) We utilize the mixture of experts technique to divide the style embedding space into multiple tractable subsets, improving the ability of the style transfer TTS model to handle more diverse and unseen speech styles.
2) We develop a gating network to route reference speech to appropriate style experts. Here, we also introduce sparsity to ensure that the computation cost at both training and inference doesn't scale with the increase in model capacity.
3)  Our method is designed to be easily applied various style transfer TTS frameworks to improve the performance of style encoders.
\vspace{-2mm}
\begin{figure*}
    \centering
    \includegraphics[width=\linewidth]{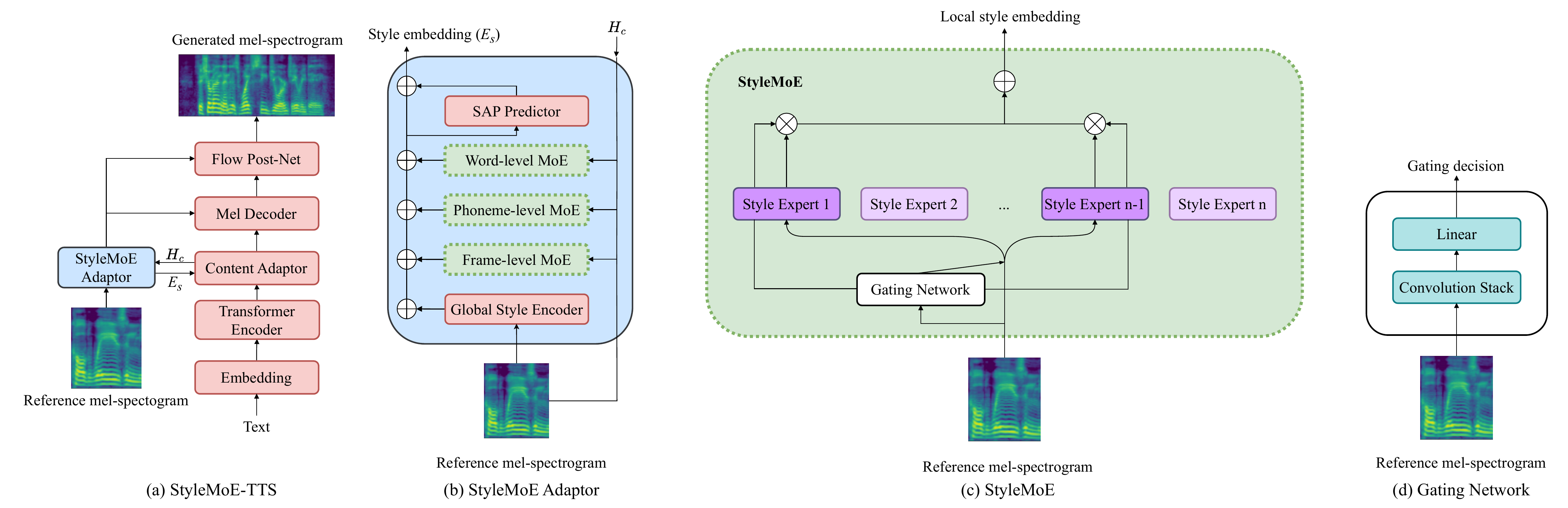}
    \caption{The  architecture of StyleMoE-TTS.
    Red modules represent modules from GenerSpeech \cite{huang2022generspeech}. Green modules represent the Mixture of Experts layer. Purple modules represent style experts. The darker purple modules represent the style experts chosen by the gating network.
    Subfigures (a) and (b) illustrate the integration of StyleMoE into StyleMoE-TTS. Subfigure (c) depicts the StyleMoE layer, wherein each style expert block is a style reference encoder. Subfigure (d) illustrates the gating network.}
    \label{fig:stylemoe}
\end{figure*}
\section{Related Work}

\subsection{Mixture of Experts}

The Mixture of Experts technique, initially introduced by Jacobs et al.~\cite{jacobs1991adaptive}, has experienced a resurgence in various domains, especially in large-scale neural networks~\cite{shazeer2017outrageously}. MoE architectures segment a problem space into subspaces, each handled by an expert. The gating mechanism partitions the space by acting as a router, directing input data to experts, causing them to specialize through the process of optimization~\cite{jacobs1991adaptive, masoudnia2014mixture}. MoE models can be categorized into two types: one that implicitly partitions the problem space through a gating network optimized by based on a loss function, and another that explicitly divides the space, often using clustering techniques to identify subspaces before training begins \cite{masoudnia2014mixture}.

In the TTS domain, our research builds upon and diverges from existing works like Teh et al.~\cite{teh2023ensemble} and Adaspeech 3~\cite{yan2021adaspeech}, which incorporated ensembles or a MoE to enhance the modeling of prosodic features, including pitch, energy, and duration. Instead of predicting individual low-level prosodic descriptors, our approach leverages MoE to directly model style representations instead.
\vspace{-2mm}
\subsection{Style Transfer in TTS}
\vspace{-3mm}
Past works have explored various conditioning techniques to overcome the challenges presented by the one-to-many mapping problem in TTS . Initial approaches focused on incorporating categorical labels, such as emotion and speaker identity \cite{7953089, lee2017emotional}, to guide the speech synthesis process. Further innovations introduced the concept of style transfer from reference speech, utilizing neural style encoders to capture the stylistic nuances (e.g., timbre, emotion, and prosody) of a given speech sample \cite{skerry2018towards, wang2018style, akuzawa2018expressive}, allowing for a more direct and effective incorporation of expressive elements into synthesized speech. For example, Meta-StyleSpeech \cite{min2021meta} proposes a style-adaptive normalization layer that aligns the gain and bias of text input with regard to extracted latent style representation, achieving high expressiveness transfer from a single short-duration speech reference.
Despite advancements, capturing a wide range of styles, especially unseen ones, remains challenging due to models' limited generalization and the complexity of modeling speech styles. To address this, we propose StyleMoE, which "divides and conquers" the style modeling problem by using specialized style experts to model subsets of data.

\section{Method}
\vspace{-3mm}
In this work, we introduce StyleMoE, a method to enhance the style coverage of a TTS model by using a MoE on the style encoder. The fundamental concept behind StyleMoE is generalizable and can potentially be incorporated into other style transfer TTS framework. StyleMoE involves replacing the original style encoder in a TTS framework with a sparse Mixture of Experts~\cite{shazeer2017outrageously}, wherein each expert in the MoE retains the original style encoder architecture but maintains separate parameters. We chose to use a sparse MoE due to its ability to implicitly partition the style space without the need for additional task-specific knowledge, which is required for explicit strategies. Furthermore, introducing sparsity enables our method to be trained without incurring extra computational costs, making it easier to integrate with existing TTS frameworks.

We implement StyleMoE on the GenerSpeech \cite{huang2022generspeech} framework (Figure \ref{fig:stylemoe} (a)) by incorporating our sparse MoE layer into each of its local style encoders. GenerSpeech uses a hierarchical style encoder, which consists of a global style encoder and local style encoders of varying resolutions. Since the global style encoder is pre-trained and remains unchanged during training, our MoE (Mixture of Experts) approach is applied exclusively to the local style encoders, as illustrated in Figure \ref{fig:stylemoe}(b). It is also important to note that the number of experts that can be utilized is limited by the parameter count of each local style encoder.

\subsection{The Mixture of Experts Layer}

Our approach utilizes the sparse MoE implementation detailed by Shazeer et al. \cite{shazeer2017outrageously} \footnote{Spare MoE Implementation: https://github.com/davidmrau/mixture-of-experts}, which consists of $n$ experts and a gating network. Each expert, denoted as $E_i$, follows the same style encoder architecture but maintains distinct parameters; hence, we call them style experts. The gating network, $G$, determines the contribution of each style expert to the final output based on input $x$, enabling a sparse, efficient computation. The local style embedding $y$ is computed as described in Equation \ref{eq1} and as illustrated in Figure \ref{fig:stylemoe}(c). \\
\begin{minipage}{0.35\textwidth} 
\begin{equation}
\label{eq1}
y = \sum_{i=1}^{n} G(x)_i E_i(x)
\end{equation}
\end{minipage}
\hspace{8mm} 
\begin{minipage}{0.50\textwidth} 
\vspace{-3mm}
\begin{equation}
\label{eq2}
G(x) = \text{Softmax}(\text{KeepTopK}(H(x), k))
\end{equation}
\end{minipage}

\subsection{Gating Network}
The gating network controls the specialization of style experts by directing different reference speech samples to certain experts. The gating network $G$ is defined as described in Equation \ref{eq2}.

The noisy top-k gating function, $H(x)$, introduces Gaussian noise to balance the selection of experts and ensure that each expert is utilized. It is described as follows:
\begin{equation}
H(x)_i = RouterNetwork(x) + \text{StandardNormal}() \cdot \text{Softplus}((x \cdot W_{noise})_i)
\label{eq:3}
\end{equation}

Here, the amount of noise per component $i$ is controlled by a second trainable weight matrix $W_{noise}$.
The RouterNetwork processes the reference speech samples and outputs an $n$-dimensional vector, as shown in Figure \ref{fig:stylemoe} (d). The network consists of a convolutional stack followed by a linear layer.

\[
\text{KeepTopK}(v, k)_i =
\left\{
\begin{array}{ll}
v_i, & \textit{if } v_i \textit{ is in the top } k \textit{ elements of } v \\
-\infty, & \text{otherwise}.
\end{array}
\right.
\]

The KeepTopK function enforces the model’s sparsity by limiting the active style experts to only the top $k$ most relevant ones. It does this by setting the weights of the less relevant experts to $-\infty$, which effectively gives them a weight of $0$  after the Softmax operation is performed.

\vspace{-3mm}
\subsection{Training and Inference}

During training, the StyleMoE adaptor and the TTS system are jointly optimized using the same learning objective as the original TTS system, which, in this case, is GenerSpeech. Our model was trained using a batch size of 32 for 300,000 steps. During inference, StyleMoE processes reference speech starting at the gating network $G$, which determines the weightings of each style expert. These weightings identify the top $k$ experts, where $k$ is adaptable at inference. The predictions from the selected style experts are then combined according to their weightings, as described in Equation (\ref{eq1}). Finally, the resulting style embeddings are used to condition the speech generation in GenerSpeech.

\vspace{-3mm}
\section{Experimental setup}
\vspace{-3mm}
\subsection{Training and evaluation data}
We trained the StyleMoE-TTS framework on the "train-clean-100" subset of the LibriTTS dataset \cite{zen2019libritts}, comprising 100 hours of multi-speaker speech data. We downsampled the dataset from 24 kHz to 16 kHz. For our evaluations, we conducted objective assessments using randomly chosen speech and texts from the Emotional Speech dataset (ESD) \cite{zhou2021seen}. We conducted our objective and subjective evaluations from samples of the ESD dataset to evaluate how well our methods perform on unseen samples that have diverse styles in speaker and emotion styles.
\vspace{-2mm}
\subsection{Baselines}
\vspace{-2mm}
We use the official unmodified implementation of GenerSpeech \footnote{GenerSpeech implementation:  https://github.com/Rongjiehuang/GenerSpeech} as one of our baselines since we implemented our method on top of GenerSpeech. 
We developed an ensembled version of the style encoder to demonstrate the specific advantage of using mixtures of experts for modeling the style space. In the ensembled version, we replaced the StyleMoE described in Figure \ref{fig:stylemoe}(c) with the StyleEnsemble encoder. The StyleEnsemble encoder, similar to StyleMoE, consists of $n$ individual style encoders, each with distinct parameters. However, unlike StyleMoE, which selects specific encoders, StyleEnsemble averages the outputs of all style encoders, creating an ensemble. This approach allows for a direct comparison to a style encoding model with a comparable parameter count to StyleMoE-TTS, trained under the same conditions. Here, we note that StyleEnsemble-TTS is a method that we developed for the purpose of comparison and it does not exist in the literature. We opted for a smaller number of experts in our experiments for two main reasons: 1) Each style encoder has around 1 million parameters, so increasing the number of experts would raise computational costs, and 2) Prior research on the application of ensemble and MoE methods in TTS has shown that fewer experts can still produce improved results. For example, AdaSpeech 3 successfully used only three MoE experts in their duration predictor \cite{yan2021adaspeech}, and ensemble methods saw improvements with just two and three models \cite{teh2023ensemble}, suggesting that a smaller number of experts is sufficient for the successful application of this method.

\vspace{-4mm}
\section{Results and Discussion}
We consider three models in our evaluation, as shown in Table \ref{table:combined}: 1) $GenerSpeech$, 2) \textit{StyleEnsemble-TTS}: GenerSpeech with an ensemble of two style adaptors, and 3) \textit{StyleMoE-TTS} ($N=2,k=1$): GenerSpeech with two style experts, with only one being used during inference.

\subsection{Objective evaluations}
\vspace{-2mm}

We evaluate speaker similarity of synthesized speech using both parallel and non-parallel text inputs by calculating the cosine distance of d-vector embeddings \cite{wan2018generalized} between synthesized and reference speech. StyleMoE-TTS shows higher speaker similarity than baselines. We also measure spectral distortion using mel-cepstral distortion (MCD) \cite{kubichek1993mel}, where lower values indicate better quality and lower distortion. StyleMoE-TTS achieves consistently lower MCD scores compared to the baselines. Finally, we report F0 frame error (FFE) \cite{chu2009reducing}, where StyleMoE-TTS achieves lower FFE, indicating better capture of low-level prosodic characteristics. We include a gating analysis in the appendix sec. \ref{sec:appendix} to visualize the  utilization of style experts across hierarchical levels and emotions.

\vspace{-2mm}
\subsection{Subjective evaluations}
\vspace{-2mm}
\begin{table}
\caption{Objective and Subjective Evaluations on ESD. Mean Opinion Score (MOS), Style Mean Opinion Score (SMOS) are reported with 95\% confidence intervals.}
\centering
\label{table:combined}
\begin{tabular}{lcccccc}
\toprule
Methods & Cos $\uparrow$ & MCD $\downarrow$ & FFE $\downarrow$ & MOS $\uparrow$ & SMOS $\uparrow$ \\
\midrule
Ground Truth & - & - & - & $4.39 \pm{0.11}$ & - \\ 
\midrule
GenerSpeech \cite{huang2022generspeech} & 0.73 & 6.00 & 0.35 & $3.66 \pm{0.13}$ & $3.46 \pm{0.12}$ \\
StyleEnsemble-TTS & 0.74 & 5.57 & 0.36 & $3.69 \pm{0.13}$ & $3.38 \pm{0.13}$ \\
\textbf{StyleMoE-TTS} ($N=2,k=1$) & \textbf{0.75} & \textbf{5.54} & \textbf{0.34} & \textbf{3.82 $\pm$ 0.11}& \textbf{3.55 $\pm$ 0.11} \\
\bottomrule
\end{tabular}
\end{table}
\begin{figure*}
    \centering
    \includegraphics[width=\linewidth]{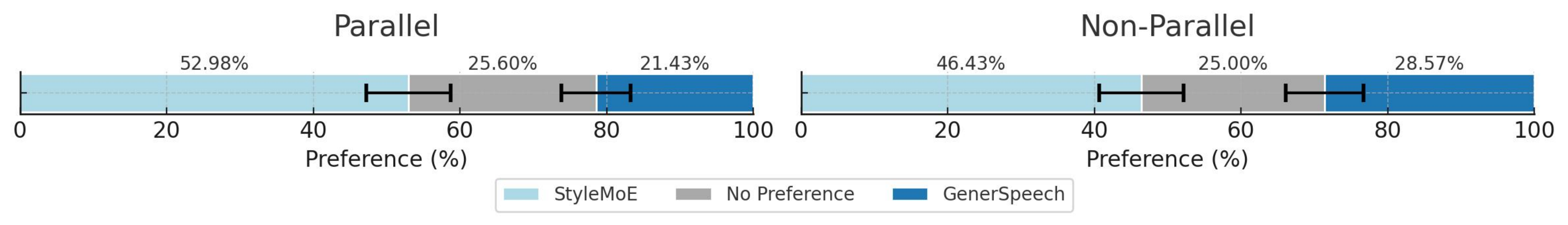}
    \caption{Style preference test on ESD reported with 95\% confidence intervals.}
    \label{fig:stylemoepref}
    \vspace{-4mm}
\end{figure*}

We conduct listening experiments with 12 subjects to evaluate the generated speech's naturalness the capability of style transfer of the proposed StyleMoE-TTS, using 144 utterances in total. For the Mean Opinion Score (MOS) evaluation \cite{streijl2016mean}, participants rate the naturalness of 12 speech samples generated by each model on a five-point scale using text from the ESD dataset. As shown in Table \ref{table:combined}, StyleMoE-TTS outperforms the baselines, achieving the highest MOS score of 3.55. The partitioned style embedding space helps the style experts learn more naturalistic styles during training.

We evaluate the models' style transfer capacity using a Style Mean Opinion Score (SMOS) test \cite{chen2024stylespeech}, where participants rate how closely the speaking style of synthesized utterances matches a provided reference speech. Ratings are on a 5-point scale, with five indicating a close match to the reference style and one indicating no similarity. We generate 16 speech samples using parallel utterances from the ESD dataset for each model. As shown in Table \ref{table:combined}, listeners consistently rate StyleMoE-TTS samples as more similar to the reference speech compared to the baselines. Thus, the sparse StyleMoE technique in StyleMoE-TTS effectively captures a wider range of speaking styles.

We conduct a style preference test \cite{li2024spontts} in which participants are presented with a reference speech sample and generated samples from StyleMoE-TTS ($N=2,k=1$) and GenerSpeech. They are asked to choose the generated sample that more closely matches the speaking style of the reference. We consider two settings: parallel, where the generated and reference speeches share the same text, and non-parallel, where they differ in content. As shown in Figure \ref{fig:stylemoepref}, in the parallel setting, StyleMoE-TTS is preferred 52.98\% of the time, while GenerSpeech is preferred 21.43\%. A similar trend is observed in the non-parallel setting, with StyleMoE-TTS being chosen 46.43\% of the time compared to GenerSpeech's 28.57\%. These results indicate that listeners prefer the proposed method over GenerSpeech, demonstrating the performance improvement of StyleMoE-TTS.

\vspace{-3mm}
\section{Conclusion}

In this work, we present StyleMoE, a method that enhances the style coverage of style encoders in style transfer TTS models. Due to the one-to-many mapping issue in expressive speech, style encoders tend to learn averaged style representations. To address this, we replaced the style encoder with a sparse mixture of experts layer consisting of style experts. Each style expert is  trained on a subset of the data, thereby mitigating the averaging problem by focusing on smaller subsets. Through our experiments, we demonstrate that our method improves style transfer performance on unseen and diverse speaking styles. Future work could explore different gating network architectures, an explicit partitioning gating strategy, and applying a hierarchical MoE.

\small
\bibliographystyle{IEEEtran}
\bibliography{neurips_2024}



\section{Appendix / supplemental material}

\subsection{Gating Analysis}
\label{sec:appendix}
\begin{figure}[!th]
    \centering
    \includegraphics[width=0.45\linewidth]{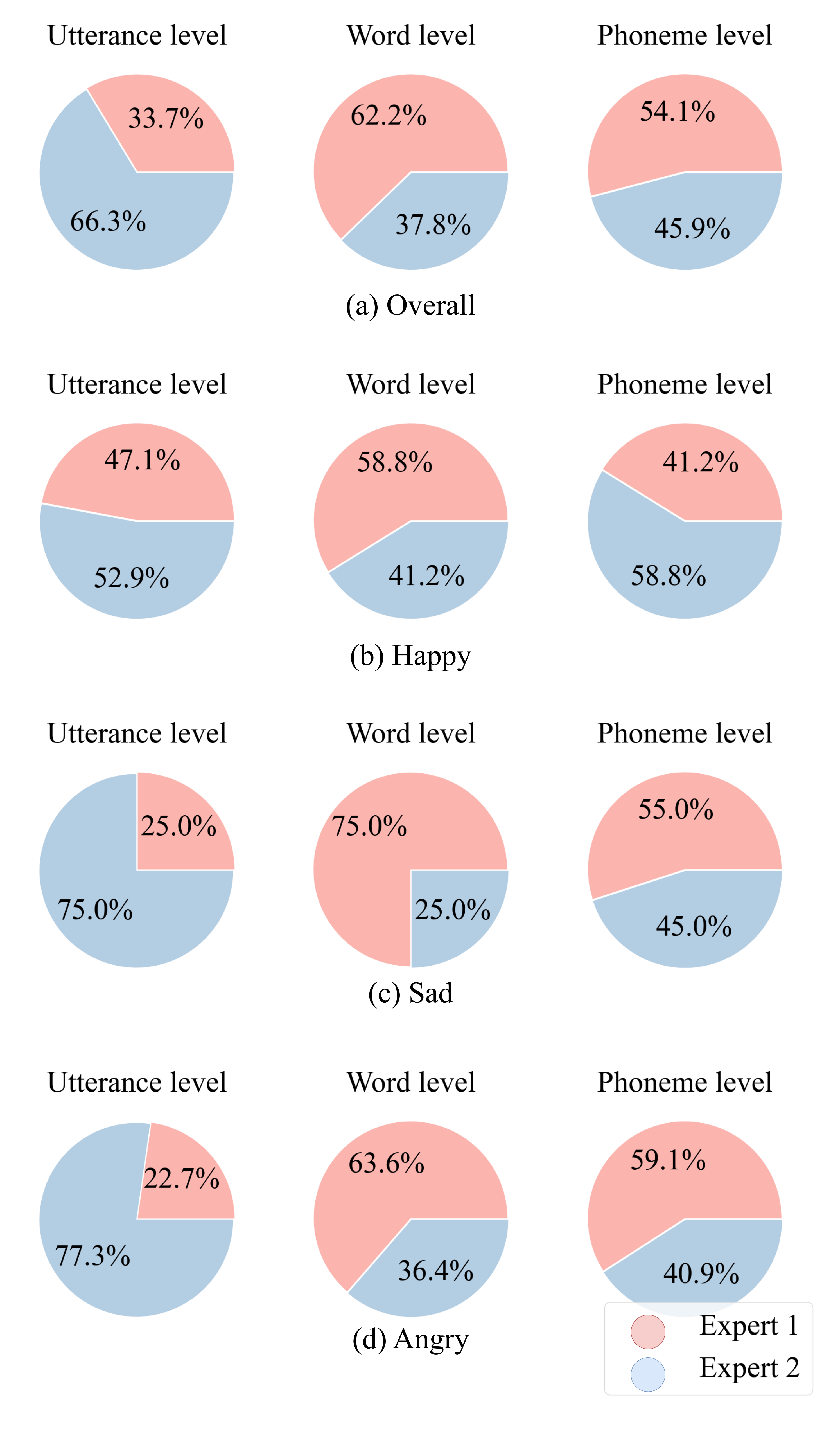}
    \caption{
    Illustration of style expert utilization in StyleMoE layers ($N=2, k=1$). Each pie chart in a row represents a separate StyleMoE Layer across different hierarchical levels. Percentages are indicative of the style expert usage. The analysis is performed over all samples in (a) and on emotion subsets in (b), (c) and (d).}
    \label{fig:gating}
\end{figure}
In Figure \ref{fig:gating}, we examine how the StyleMoE's gating network routes samples across different experts, highlighting the division of the style embedding space. It is important to note that our gating network, which partitions data samples implicitly rather than using a clustering algorithm, encourages a distributed combining strategy.

The style adapter consists of three local style encoders which differ in the resolution of style encoding - utterance level, word level and phoneme level. Each of the local style encoders is modeled as a mixture of $N$ experts. For this experiment, we use a two-expert ($N=2$) framework, where a single expert ($k=1$) is selected based on the input reference speech. We analyze 100 utterances that are uniformly chosen across the emotion categories from the ESD dataset.
We observe the following:
\begin{enumerate}
    \item Both style expert one and style expert two are utilized, indicating that the task of style modeling is shared among all the style experts in the style adapter.
    \item In the row-wise comparison of the pie-charts in Figure \ref{fig:gating}(a), (b), (c) and (d), we observe the experts are utilized differently across hierarchical levels. From Figure \ref{fig:gating}(a), we see that style expert one is utilized more at word level, while style expert two is utilized more at utterance level. This suggests that the experts are learning different information at varying resolutions, indicating that the proposed "divide and conquer" strategy may have been effectively applied.
    \item Figures \ref{fig:gating} (b), (c), and (d) depict variations in expert utilization based on emotion. Here, we observe the experts are being utilized differently with different emotions, which indicates that the gating network learns a routing strategy that is based on the distribution of speaking styles. 
\end{enumerate}

\end{document}